\documentclass[twocolumn,showpacs,preprintnumbers,amsmath,amssymb,floatfix]{revtex4}

\usepackage{graphicx,epsfig}
\usepackage{dcolumn}
\usepackage{bm}

\def\pllc{$K^+ \to \pi^+ l^+ l^-$}
\def\plll{$K^0_L \to \pi^0 l^+ l^-$}

\def\pmml{$K^0_L \to \pi^0 \mu^+ \mu^-$}

\begin{document}
\title{ 
  Muon Decay Asymmetries from  $K^0_L \to 
\pi^0 \mu^+ \mu^-$ Decays
  }
\author{
Milind V. Diwan, Hong Ma, T. L. Trueman}
\affiliation{ 
 Physics Department,
Brookhaven National Laboratory, Upton, NY 11973
}

\begin{abstract}
  We have examined
the decay
$K^0_L \to \pi^0 \mu^+ \mu^-$
 in which  the branching ratio, the muon energy
asymmetry and the muon decay asymmetry
 could be  measured.
In particular, we find that
within the Standard Model
 the longitudinal polarization ($P_L$) of the muon
is proportional to  the direct CP violating amplitude.
On the other hand the energy asymmetry and the out-of-plane
polarization ($P_N$) depend on both indirect and direct CP violating
amplitudes.
Although the branching ratio is small and difficult to
measure because of background, the asymmetries could 
be large $\cal{O}$(1)
in the Standard Model.
 A combined analysis of
the  energy asymmetry, $P_L$ and $P_N$
 could be used to separate indirect CPV, direct CPV, 
and CP conserving
contributions to the decay.
\end{abstract}


\maketitle

\section{Introduction}

\noindent

 There are three possible
contributions to the \plll ~decay amplitude: 1) direct
CP-violating contribution from electroweak penguin
and W-box diagrams, 2) indirect CP-violating amplitude
from the $K_1 \to \pi^0 l^+ l^-$ component in $K_L$, and
3) CP-conserving amplitude from the $\pi^0 \gamma \gamma$
intermediate state {\cite{winwolf,rwoj}}.
 The sizes of the three contributions
depend on whether the final-state lepton is an electron or a muon.
The CP-conserving two-photon contribution to the electron 
mode is expected to
be $(1-4)\times10^{-12}$, based on $K_L\rightarrow \pi^0 
\gamma \gamma$ data.
Although suppressed in phase space, this contribution 
to the muon mode is
comparable to the electron mode because of the scalar form factor
which is  proportional to
lepton mass\cite{hsehgal, epr291}.
The interesting direct CP-violating component
must be extracted from any signal found
for $K_L \to \pi^0 l^+ l^-$ in the presence of two
 formidable obstacles:
 the theoretical uncertainty on contamination from indirect
 CP-violating and
CP-conserving contributions, and
 the experimental background
from  $l^+ l^- \gamma \gamma$ \cite{greenlee}.

To subtract the indirect CP-violating and CP 
conserving contributions,
several authors have examined
the use of measurements from
$K_L \to \pi^0 \gamma \gamma$, $K_S \to \pi^0 e^+ e^-$, as well as
 the lepton energy
asymmetry{\cite{rwoj,donoghue,dambrosio,bgeng91,gabval}}.
With better measurements expected in the near future, it is useful
to reexamine \plll decays 
\cite{piggktev, na48ks}.

In this paper, we examine if the muon decay asymmetries
 give additional
constraints on CP violation in \pmml.   Indeed, we find that
within the Standard Model the P-odd longitudinal polarization
of the muon is non-zero only if the direct CP violating
amplitudes are non-zero. We also find that other asymmetries
that involve the polarization of both muons can be constructed to
isolate the direct and indirect contributions to CP violation.
In sections \ref{matel} and \ref{secpol} we discuss the phenomenology of 
the decay and polarization. In section \ref{formfac} we 
describe the form factors used for our numerical 
estimates. In section \ref{numres} we discuss the numerical
results. We have computed the results using, as much as possible, 
previously obtained results on various form factors; we also discuss 
future necessary theoretical calculations for reducing the
large  uncertainties on our estimates. We conclude in  section \ref{conclud}
with a description of the current experimental situation and future 
possibilities.

\section{Matrix Elements}
\label{matel}

We will proceed in analogy to the charged version of the decay,
$K^+\to \pi^+ \mu^+ \mu^-$
which has been analyzed extensively
\cite{anbg, wise1, wise2, buchalla, gourdin, bgt}. 
The structure of
 $K_L\to \pi^0 l^+ l^-$ decays is more complex than
 that of  $K^+ \rightarrow \pi^+ l^+ l^-$ decays
because of  CP suppression.
In particular,
 the one-photon intermediate state contribution to the
vector form factor ($F_V$), which is expected to be almost
real and dominant for \pllc ~decays, is
CP-suppressed in \plll ~decays.

The \plll decay process can have contributions from
scalar, vector, pseudo-scalar and
axial-vector interactions, with corresponding form factors, 
$F_S$, $F_V$,
$F_P$, and $F_A${\cite{anbg}}:
\begin{eqnarray}
{\cal M} & = & F_S \bar u(p_l, s)v(\bar p_l, \bar s)
+ F_V p^\mu_k \bar u(p_l, s)\gamma_\mu  v(\bar p_l, \bar s) 
\nonumber \\
 &  + & F_P \bar u(p_l, s)\gamma_5 v(\bar p_l, \bar s)  +
    F_A p^\mu_k \bar u(p_l, s)\gamma_\mu  \gamma_5 v(\bar p_l, 
\bar s) 
\nonumber  \\
& & 
\label{amppmm}
\end{eqnarray}
Here $p_k$, $p_l$, and $\bar p_l$ are the kaon, 
 lepton, and
antilepton 4-momenta.
Transverse muon polarization effects arise from 
interference between terms
with non-zero phase differences, while longitudinal 
polarization results
from interference between terms with the same phase.

 The scalar form factor, $F_S$,  is expected to get
a  contribution from only the two-photon intermediate state,
$K_L \to \pi^0 \gamma \gamma \to \pi^0 \mu^+ \mu^-$.
Pseudo-scalar, $F_P$,  and  axial-vector, $F_A$, get 
contributions from the
short-distance ``Z-penguin'' and ``W-box'' diagrams only,
 where the dominant term arises
from t-quark exchange;  both form factors therefore  depend on
$V_{ts}V_{td}^*$, with a small contribution from the charm quark.
$F_V=F_V^+ + F_V^-$ where $F_V^+$ is the  CP-even
 contribution   from
the two photon process and is proportional to
$p_k\cdot(p_l-\bar{p_l})$. $F_V^-$ is the CP-odd contribution, 
the sum of
$F_V^{MM}$, the indirect CP violation in $K_L$ decays, and
 $F_V^{dir}$, the
short distance, direct CP violating contribution.
Unlike $K^+$,
all the amplitudes in the $K_L$ decay
 are likely to be of  the same order of magnitude, and
 polarization effects should therefore be  large 
(${\cal O}(1)$), unless
there are strong cancellations \cite{mpla}.

The symmetries of the decay are clearest
 in the $l^-l^{+}$ CM frame. In this reference frame,
$p_l=(E,\vec{p}), \,\bar{p_l}=(E,-\vec{p})$ and $\lambda ,\, 
\mu$ are the
helicities of the lepton
and anti-lepton, respectively.
Using this notation,  the helicity amplitudes, 
$M_{\mu \lambda}(\hat{p})$,
 in terms of the four
form factors
$F_S,F_P,F_V, F_A$ are:
\begin{eqnarray}
M_{++}(\hat{p})&=& -F_S \frac{p}{m}+F_P \frac{E}{m} - F_V\, p_k
\cos{\theta} +F_A E_k \nonumber \\
M_{--}(\hat{p}) &=& +F_S \frac{p}{m}+F_P \frac{E}{m} + F_V\, p_k
\cos{\theta} +F_A E_k \nonumber \\
M_{+-}(\hat{p}) &=& +F_V \frac{E}{m}\, p_k\, \sin{\theta}  -F_A
\frac{p}{m}\, p_k
\,\sin{\theta} \nonumber \\
M_{-+}(\hat{p}) &=& +F_V \frac{E}{m}\, p_k \,\sin{\theta} +F_A
\frac{p}{m}\, p_k
\,\sin{\theta}
\end{eqnarray}
Where $p$, $E$, $m$ are the momentum, energy and the mass of the
negative lepton, and $\theta$ is the angle between the 
negative lepton
and the kaon in the $l^-l^+$ CM frame.
$M_{\mu,\lambda}(\hat{p})$ can depend on various invariants like
$p_k\cdot (p_l+\bar{p_l})$ but
we single out  the unit vector 
$\hat{p}=\vec{p}\,/|\,\vec{p}\,|$ because it
changes under the $CP$-operation.
Under $CP$ $M_{\mu,\lambda}(\hat{p}) \rightarrow
-M_{-\lambda,-\mu}(-\hat{p})$. As we have already discussed,
$F_V = F_V^+ + F_V^-$
and $F_V^{\pm}(\hat{p}) = \mp F_V^{\pm}(-\hat{p})$.
 In this set of amplitudes
the form factors have no
$\theta$ dependence other than an explicit factor of 
$\cos{\theta}$ in
$F_V^+$.
These helicity amplitudes can be used to create three 
interesting CP-odd
observables shown in Eq. \ref{3guys}.  The difference 
between  the last
 two asymmetries in Eq. \ref{3guys} isolates
 the interference between
 the $CP=+1$ part of $F_V$  and purely direct CP 
violating amplitudes.
These asymmetries involve the measurement of 
the polarization of both
leptons in the same event. 
\begin{widetext}
\begin{eqnarray}
|M_{++}(\hat{p})|^2 - |M_{--}(-\hat{p})|^2 & = &  -4 
Re\{(F_S \frac{p}{m} +
F_V^+p_k
\cos{\theta})^* \times (F_P\frac{E}{m}+F_A E_k -F_V^- p_k 
\cos{\theta})\}  \nonumber \\
|M_{+-}(\hat{p})|^2 -|M_{+-}(-\hat{p})|^2 &=& + 4 Re\{(F_V^+
\frac{E}{m})^*  \times  
 (F_V^-\frac{E}{m} -F_A \frac{p}{m})\} p_k^2 \sin^2{\theta}
\nonumber \\
|M_{-+}(\hat{p})|^2 -|M_{-+}(-\hat{p})|^2 &=& 4 
Re\{(F_V^+ \frac{E}{m})^*
 \times    (F_V^-\frac{E}{m} +F_A \frac{p}{m})\} p_k^2 
\sin^2{\theta} 
\label{3guys}
\end{eqnarray}
\end{widetext}

It is possible to get direct
information about CP
violating amplitudes from measuring asymmetries for 
only one of the leptons.
It is
well-known{\cite{pi0mumu}} that the
out-of-plane polarization of the muon, transverse to the 
plane formed by the
$\pi$ and the
lepton momenta, gets contributions from CP-violating amplitudes,  
but several
effects are
mixed up and it is not possible to give a separation 
of the direct and the
indirect pieces. It is, however,  
not well-known that more information can be
obtained from the
 parity-violating single lepton 
longitudinal asymmetries, even
though they are not
intrinsically CP-violating. The longitudinal polarizations
 are given in the
lepton-lepton center of mass  frame in Eq. \ref{thems}.
We use the notation $P'_L$ for 
the polarization components
measured in the
$l \bar{l}$ cm to distinguish them from the  $P_L$ measured in
the kaon rest
frame, to be used later.
Longitudinal polarization is not an indicator of $CP$ violation
as it is of $P$ violation; $CP$ conservation only requires 
that $F_P$, $F_V$, and $F_A$ be odd under 
$\hat{p} \rightarrow -\hat{p}$, not that they vanish. For the 
amplitudes  considered here, this is not so, and since every term
in Eq. \ref{thems} has either $F_P$ or $F_A$ as a factor, 
$P_L'(\hat{p})$ is a direct measure of $CP$ violation in 
the decay. Furthermore, unlike the polarization normal to 
the decay plane, a common indicator of $T$-violation, it can
be used to separate the direct from the indirect 
$CP$ violation. 
\begin{widetext}
\begin{eqnarray}
P'_L(\hat{p}) &=& \left[|M_{++}|^2 +|M_{-+}|^2 - 
|M_{+-}|^2 -|M_{--}|^2
\right] /\rho \nonumber \\
    &=& [-4 Re\{(F_S\frac{p}{m} +F_V p_k 
\cos{\theta})^*(F_P \frac{E}{m} +
F_A E_k)\}   + 4 \frac{pE}{m^2}\, p^2_k
\sin^2{\theta} Re(F_A^*F_V) ]/\rho, \nonumber \\
\bar{P'}_L(-\hat{p}) &=& \left[|M_{++}|^2 +|M_{+-}|^2 - 
|M_{-+}|^2
-|M_{--}|^2\right]/\rho
\nonumber \\
    &=& [-4 Re\{(F_S \frac{p}{m} +F_Vp_k 
\cos{\theta})^*(F_P \frac{E}{m} +
F_A E_k)\}   - 4 \frac{pE}{m^2}\, p^2_k
\sin^2{\theta} Re(F_A^*F_V) ] /\rho, \nonumber \\
\rho & = &  |M_{++}|^2 +|M_{-+}|^2 + |M_{+-}|^2 + |M_{--}|^2.
\label{thems}
\end{eqnarray}
\end{widetext}
The combination
($P'_{L}(\hat{p}) + P'_{L}(-\hat{p})+\bar{P'}_L(\hat{p}) +
\bar{P'}_L(-\hat{p})$) cancels the interference between 
pairs of CP violating
amplitudes and leaves a pure CP-violating observable. Similarly
($P'_{L}(\hat{p}) + P'_{L}(-\hat{p})-\bar{P'}_L(\hat{p})-
\bar{P'}_L(-\hat{p})$) is CP-even.
The angular dependence  can be integrated to give a particularly
simple result:
\begin{eqnarray}
<P'_L + \bar{P'}_L >& = & -16 Re\{(F_S \frac{p}{m}
+\frac{1}{3}F_V^+(\theta=0)\,p_k )^*  \nonumber  \\
 & \times & (F_P
\frac{E}{m} + F_A E_k)\}/\rho \nonumber \\
<P'_L - \bar{P'}_L> & = &
\frac{32}{3}\frac{pE}{m^2}\, p^2_k
 Re(F_A^*F_V^-)/\rho
\end{eqnarray}
The first of these is C-even and purely direct, and the second is
C-odd and contains both direct and indirect amplitudes.
It should be noted that for these asymmetries
 the muon and anti-muon polarizations can be measured separately over
the same part of phase space. Indeed, if a complete 
angular analysis of one
lepton's
polarization can be performed, 4 out of the 6 
independent products of 
the
form factors can be
determined.

\section{Polarization}
\label{secpol}

We will show our numerical results in the kaon rest frame.
We concentrate on four measurable quantities: the total decay rate,
the energy asymmetry between the muon and the anti-muon, 
 the out-of-plane,
 and the longitudinal
components of the muon polarization. The in-plane 
transverse polarization
should also be considered when designing an experiment.

For the purpose of discussing possible experiments, 
it is useful to have
the lepton
polarization given in a covariant way 
as in Eq. \ref{covar}. 
Here,
$q^2=(p_l + \bar{p}_l)^2$, and $s$ denotes the covariant spin vector
of the lepton.
The decay rate in the kaon rest frame is given in Eq. \ref{drate}
 \cite{anbg}.

\begin{widetext}
\begin{eqnarray}
P(s) & = & \{-2 Re(F_SF_P^*) m (s\cdot \bar{p}_l) \nonumber 
+ 2 Re(F_VF_A^*) m [2(s\cdot p_k)(\bar{p}_l \cdot p_k) - m_K^2 
(s \cdot
\bar{p}_l)] \nonumber \\
&-& 2Re(F_PF_V^*) [ -\frac{1}{2} q^2(p_k \cdot s) + 
(p_k \cdot p_l)(\bar
{p}_l \cdot
s)] + 2 Re(F_S F_A^*)[\frac{1}{2}(q^2 - 4 m^2)(p_k \cdot s) 
-(\bar{p}_l \cdot
s)(p_l \cdot
 p_k)]\nonumber \\ &+& [2 Im(F_PF_A^*) + 2 Im(F_SF_V^*)]
 \epsilon^{\mu \nu
\rho \sigma}p_{k\mu}
p_{l\nu} \bar{p}_{l\rho} s_{\sigma}\}/(m^2 \rho) \,,
\label{covar}
\end{eqnarray}
\begin{eqnarray}
m^2 \rho_0(E_l, \bar{E}_l) &=&  |F_S|^2\frac{1}{2}(q^2-4m^2)
+|F_P|^2\frac{1}{2}q^2 
     + |F_V|^2 m_k^2 (2E_l\bar{E}_l - \frac{1}{2}q^2) \nonumber \\
 &+& |F_A|^2 m_k^2 [ 2E_l\bar{E}_l - \frac{1}{2}(q^2-4m^2)] 
+ 2Re (F_SF_V^*)m m_k (\bar{E}_l-E_l) \nonumber \\
 &+& 2Re(F_PF_A^*)\frac{m}{2}(m_k^2-M_\pi^2+q^2)].
\label{drate}
\end{eqnarray}
\end{widetext}

The total decay rate is given by
\begin{eqnarray}
 \Gamma = \frac{m^2}{2 m_k}\int \rho_0(E_l,\bar{E}_l)
\frac{dE_ld\bar{E}_l}{(2\pi)^3}.
\end{eqnarray}

The spin vector $s_L$ in the direction of the $\mu^-$ 
momentum in the kaon
rest frame is
\begin{equation}
s_L = (p, E_l \sin{\theta_{l \bar{l}}}, 0, E_l 
\cos{\theta_{l \bar{l}}})/m
\end{equation}
where we take the decay to be in the $x-z$ plane 
with $\vec {\bar{p}}_l$
pointing in the
$z$-direction. Then the longitudinal polarization of $\mu^-$ 
in the kaon rest frame
(not the same as Eq. \ref{thems}
which is in the $\mu^-
\mu^+$ center of mass) is given as
\def\vp{\vec{p}}
\def\pp{\vp_l\cdot\vec {\bar{p}}_l)}
\begin{eqnarray}
 P_L & = &  [- 2 Re (F_SF_P^*) (\bar{E}_l - \frac{E_l}{p_l^2} \pp )
\nonumber \\
    & + & 2 Re(F_VF_A^*)m_k^2(\bar{E}_l+\frac{E_l}{p^2_l}\pp)
 \nonumber  \\
    & + & 2Re(F_PF_V^*)m_k(m+\frac{m}{p_l^2}\pp) \nonumber  \\
    & + & 2Re(F_SF_A^*)m_k(-m+\frac{m}{p^2_l}\pp) ]\, 
\frac{p_l}{(m^2 \rho_0)}
\end{eqnarray}
where $\vp_l(\vec {\bar{p}}_l)$ and $E_l(\bar{E}_L)$ 
are the momentum and
energy of
$\mu^-(\mu^+)$. Notice that in this frame $F_V^+$ is odd under
$E_l\leftrightarrow  E_{\bar l}$,
while
 $F_V^-$ and the other form factors are even under 
the same interchange.

The transverse (out of decay plane) polarization 
perpendicular to the muon
momentum vector in the kaon center of mass frame is given as
\begin{eqnarray}
P_N & =&  2 [Im(F_S F_V^*)+Im(F_P F_A^*)]\,p_l \, 
{\frac {{\bar{p}}_l\, \sin{\theta_{l
\bar{l}}}}  {(m^2\rho_0)}}.
\end{eqnarray}
It should be noted that  $F_P$ and $F_A$ are in 
phase and therefore do not
contribute to $P_N$.

For completeness,  we also give  the expression for the transverse,
in-the-plane polarization $P_T$. For this the spin vector is  $s_T
= (0, \cos{\theta_{l
\bar{l}}}, 0 , -\sin{\theta_{l \bar{l}}})$, and 
the expression is as follows:
\begin{eqnarray}
P_T &= &-[2 Re(F_SF_P^*)\, m\, \nonumber \\ &+& 2 Re(F_P
F_V^*)\, m_K \,E_l \,
 \nonumber \\ &+& 2 Re(F_SF_A^*)\, m_K\,E_l\,
 \nonumber \\ &+& 2 Re(F_V F_A^*)\, m \,m_K^2 \,
] \,\bar{p}_l \,\sin{\theta_{l \bar{l}}}/(m^2 \rho_0).
\end{eqnarray}
It depends on the same quantities as $P_L$ and, depending on the
experimental configuration,
the measured quantity may be a linear combination of the two.
 ($P'_L$ is
the linear
combination of $P_L$ and $P_T$ given by the rotation of 
the lepton spin
through the angle
between the kaon and the lepton momenta as seen in the rest
frame of the
lepton.)

\section{Form Factors}
\label{formfac}

We now consider   the form-factors that we will use in estimating
the polarization.

\noindent
$F_V^{MM}$:
The decays $K_S\to \pi^0 l^+ l^-$  have been studied 
in chiral perturbation
theory extensively.
The same framework of analysis is often applied to the similar decay
 $K^+ \to \pi^+ l^+ l^-$.  In D'Ambrosio {\it et
al}\cite{dambrosio}, the decays are
analyzed beyond the leading order ${\cal O}(p^4)$.
The vector form factor for $K_S$ decays is parametrized as
\begin{eqnarray}
F_V^S = -G_F\frac{\alpha}{4\pi} (a_S+b_S z )
      -\frac{\alpha}{4\pi m_k^2} W^{\pi\pi}(z)
\end{eqnarray}
The function $W^{\pi\pi}$ comes from the pion loop 
contribution, which is
estimated to
${\cal O}(p^6)$ using $K\to\pi\pi\pi$ data, and the rest
 of the contributions
are parametrized in the linear term.
Using  the vector meson dominance model
 a further assumption is sometimes made,
\begin{eqnarray}
b_S & = & \frac{a_S}{M_V^2/m_k^2} = \frac{a_S}{2.5}
\end{eqnarray}
where $M_V$ is the vector meson mass.
The vector form factor for indirect CP violation in 
$K_L \to \pi^0 \mu^+
\mu^-$
 is then
\begin{eqnarray}
F_V^{MM} & = & \epsilon F_V^S  \label{eqfvm}
\end{eqnarray}
The value of $a_S$ is  unknown, however it is considered 
to be ${\cal
O}(1)$.
The study of the similar
decay $K^+ \to \pi^+ l^+ l^-$  gives a value of the 
equivalent parameter to
be $-0.59\pm0.01$,\cite{e865pee,e787pmm, e865pmm} 
therefore we use the  value, $-0.6$, for our numerical
results.

\noindent
$F_V^{dir}$, $F_A$, and $F_P$:
These form factors get contributions from short distance
box and penguin diagrams.
We use the notation from Donoghue and Gabbiani\cite{donoghue}:
\begin{eqnarray}
F^{dir}_V & = & 2\frac{G_F}{\sqrt 2}\, i\, y_{7V} 
Im{\lambda_t} \nonumber  \\
F_A & = & -2\frac{G_F}{\sqrt 2} \,i \,y_{7A} Im{\lambda_t} 
\nonumber  \\
  y_{7V} & = & 0.743 \, \alpha,\mbox{at} \, m_t=175  GeV 
\nonumber \\
   y_{7A} & = & -0.736 \, \alpha, \mbox{at} \, m_t=175  
GeV \nonumber
\nonumber \\
 \lambda_t & = & V_{td}V_{ts}^*.
\end{eqnarray}
We have used $\mbox{Im}\lambda_t = 10^{-4}$ in our 
numerical calculation. 
Similar to the case of $K^+$ decay  $F_P$ is related to
$F_A$ by\cite{anbg}
\begin{eqnarray}
F_P =-m_l (1-\frac{f_-}{f_+}) F_A \label{eqfp}
\end{eqnarray}
where $f_+$ and $f_-$ are charged current semileptonic decay
form factors of $K_L$.

\noindent
$F_S$: Since \cite{donoghue} is concerned only with 
electron final states,
the scalar
contribution is negligible and they do not calculate it. 
Therefore we use
the earlier result of
Ecker {\it et al}{\cite{pi0mumu}}. They have 
calculated $F_S$ to ${\cal
O}(p^4)$ as
\begin{eqnarray}
F_S &=&\frac{iG_8\alpha^2}{4\pi}m\,E(z)\nonumber\\
E(z) &=& \frac{1}{\beta z}log(\frac{1-\beta}{1+\beta}) \nonumber \\
  & & \times   [(z-r_\pi^2)F(z/r_\pi^2) - (z-1-r_\pi^2)F(z)].
 \label{eqfs}
\end{eqnarray}
$z= (p_l + \bar{p_l})^2/m_k^2$, $r_\pi = m_\pi/m_k$, and
$\beta = \sqrt{1-4\frac{m_l^2}{m_k^2 z}}$. $F(z)$ is a known function
described by Ecker {\it et al}.

\noindent
$F_V^+$: For this we return to \cite{donoghue}.
 Using chiral perturbation theory they obtain
\begin{eqnarray}
F_V^+ &=& 2  G_8\alpha^2 p_k\cdot(p_l-\bar{p_l}) K(s) \nonumber \\
 K(s) &=& \frac{B(x)}{16\pi^2m_k^2}[
 \frac{2}{3}ln(\frac{M_\rho^2}{-s})
          -\frac{1}{4}ln(\frac{-s}{m^2})+\frac{7}{18}] \label{eqfvp}
\end{eqnarray}
where $s=(p_l+ \bar{p_l})^2$, $x=s/4M_\pi^2$, and
$B(x)$ is a complex function described in \cite{donoghue}. 
$G_8$ denotes
the octet coupling
constant in chiral perturbation theory.  We use the value $G_8 =
\frac{G_F}{\sqrt{2}}
|V_{ud}V_{us}^*| g_8$ with $g^{tree}_8 = 5.1$\cite{donoghue}.
 The value
of $g^{loop}_8=
4.3$ from a higher order calculation does not alter  our conclusions
significantly \cite{g8}. The function  contains  a free parameter,
$a_V = -0.72\pm 0.08$, determined from
experimental $K_L\to \pi^0 \gamma \gamma$ data\cite{piggktev}.
We have used $a_V = -0.70$ for our numerical results.
It should be noted that Heiliger {\it et al} \cite{hsehgal}
have performed an analysis of both $F_S$ and $F_V^+$ in 
a two component
(pion loop and vector meson dominance) model.
This calculation, however, needs to be checked against the
latest data on $K_L \to \pi^0 \gamma \gamma$. 
Recently, Gabbiani and Valencia \cite{gabval} have pointed out that a more complete
formulation of
$F_V^+$, from  Cohen, Ecker, and Pich \cite{cohen},
requires three free parameters which can be obtained from
$K_L \to \pi^0 \gamma \gamma$ data.

\section{Numerical Results}
\label{numres} 

\begin{figure}
\begin{center}
\epsfig{figure=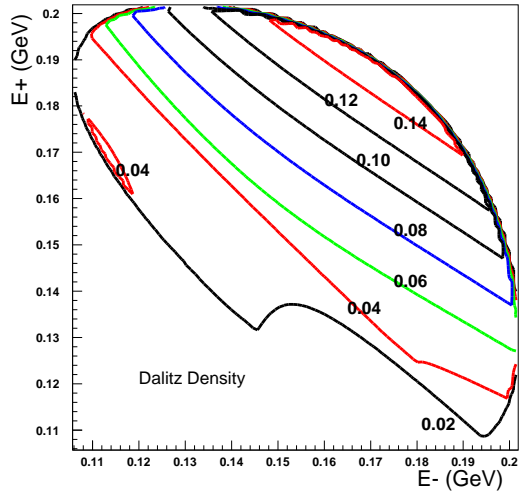,height=3in,width=3in}
\end{center}
\caption{Dalitz decay distribution 
for $K_L \to \pi^0 \mu^+ \mu^-$.
 The units for the decay rate contours
 are arbitrary. The total
calculated branching fraction
was $6.6\times 10^{-12 }$. The form factors and the 
values of the parameters used are described
in the text.   }
\label{dalitz}
\end{figure}

\begin{figure}
\begin{center}
\epsfig{figure=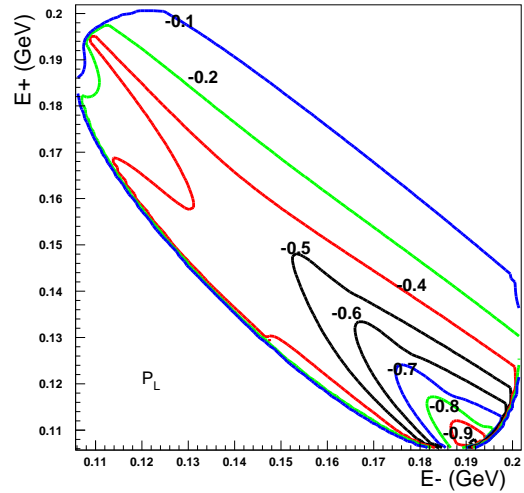,height=3in,width=3in}
\end{center}
\caption{Longitudinal ($P_L$)  polarization  of
$\mu^+$ in
$K_L \to \pi^0 \mu^+ \mu^-$ decay
plotted as a function of $\mu^+$ and $\mu^-$ energy 
 in the kaon rest frame. }
\label{poll}
\end{figure}

\begin{figure}
\begin{center}
\epsfig{figure=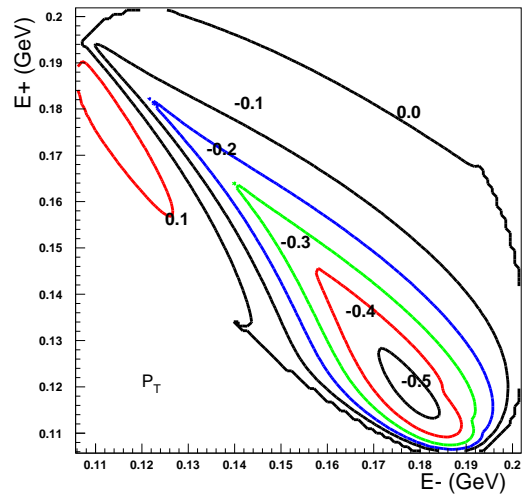,height=3in,width=3in}
\end{center}
\caption{In-plane transverse ($P_T$)  polarization  of
$\mu^+$ in
$K_L \to \pi^0 \mu^+ \mu^-$ decay
plotted as a function of $\mu^+$ and $\mu^-$ energy 
 in the kaon rest frame. }
\label{polt}
\end{figure}

\begin{figure}
\begin{center}
\epsfig{figure=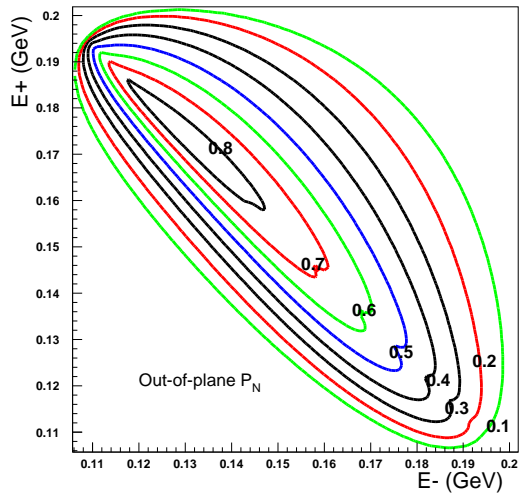,height=3in,width=3in}
\end{center}
\caption{Out-of-plane ($P_N$)  polarization  of
$\mu^+$ in
$K_L \to \pi^0 \mu^+ \mu^-$ decay
plotted as a function of $\mu^+$ and $\mu^-$ energy 
 in the kaon rest frame. }
\label{poln}
\end{figure}

Using the above discussed values for the various form factors
the total branching fraction for $K_L \to \pi^0 \mu^+ \mu^-$ 
 is about
$\sim 6.6 \times 10^{-12}$, and it
is dominated by the scalar interaction which makes most 
of the contribution
for mu-mu mass above 280 MeV.
 This is because of the two pion loop contribution
embodied in the form factor $F_S$.
Unlike  $F_S$, the rest of the form factors are expected
to give contributions that fall with $q^2$.
The contribution from $F_P$ is small because of the 
suppression due to the
lepton mass.
The region  $\sqrt{q^2} < 280$ MeV will be affected
by interference effects as shown in figure \ref{dalitz}.
It is interesting to note that, using the above described 
form factors and
parameters,  
the destructive interference causes
the decay rate to be almost zero in the region where 
the $\mu^+$ is at
rest in the kaon rest frame.
The longitudinal, in-plane transverse, and the out-of-plane
  polarizations of the $\mu^+$ are shown
in figures \ref{poll}, \ref{polt}, and \ref{poln}.
The energy asymmetry and the  polarizations are large, but they
also have large dependence on
the parameter $a_S$ and $F_S$. 
 The ${\cal O}(p^4)$ calculation for $F_S$ that we have used 
is likely to be inadequate. 
Calculations of ${\cal O}(p^6)$ contributions are beyond the
scope of this paper. Nevertheless, we have used the results 
of Cohen et al. \cite{cohen} to estimate the size of these 
contributions: they are large, but uncertain because the largest comes
from uncertain subtraction constants. With reasonable estimates
for the largest of these the change is more than a factor of 2  
over all of phase space\cite{piggktev}.  Since this is a partial, 
2 $\gamma$ unitarity cut calculation, we don't report the 
estimates here, but leave it to a more complete calculation which
must be carried out to extract important parameters from the
data.

There could be a number of ways of constraining $F_S$ as well as 
$a_S$. The branching ratio above some cut on $\mu\mu$ mass could be 
used to  fix $F_S$. 
The new expected measurement of $K_S \to \pi^0 e^+ e^-$\cite{na48ks}
will lead to constraints on the  magnitude of $a_S$. 
The out-of-plane polarization could be used to fix the sign as well
as magnitude of $a_S$, and
the short distance physics could be extracted by fitting
the measured  distribution of the longitudinal polarization which
mainly comes from $Re(F_SF_A^*)$ and $Re(F_AF_V^*)$. 

\section{Conclusion}
\label{conclud}

The modes $K_L \to \pi^0 e^+ e^-$ and $K_L \to \pi^0 \mu^+ \mu^-$
  have not as yet been  observed;
the  current  best limits on the branching ratios  
for $K_L \to \pi^0 l^+ l^-$
 were obtained by the  KTeV experiment
at FNAL;
$B(K_L \to \pi^0 \mu^+ \mu^-) < 3.8 \times 10^{-10}$
and $B(K_L \to \pi^0 e^+ e^-) < 5.1\times 10^{-10}$ 
\cite{ktevpmm, ktevpee}. 
These limits were based on  2 observed events in each case,
 and expected backgrounds  of
$0.87\pm 0.15$ for the muon mode and $1.06\pm 0.41$ for
 the electron mode.
The main backgrounds for the muon mode
 were estimated to be from $\mu^+ \mu^- \gamma \gamma$
  and $\pi^+ \pi^- \pi^0$, 
in which both
charged pions
decay in flight.  Of these, the former background could 
be irreducible and
therefore of
great concern.
For any future experiment  it seems unlikely that the
background due to $\mu^+ \mu^- \gamma \gamma$
can be lowered.
The signal to background ratio, assuming
that only $\mu^+ \mu^- \gamma \gamma$ will contribute 
in a future experiment,
will  be around 1/4, if the standard-model signal is taken as
$B(K_L \to \pi^0 \mu^+ \mu^-) \sim 6.6\times 10^{-12}$.\,{\cite{mpla}}

Measuring the muon polarization asymmetries in 
$K_L \to \pi^0 \mu^+ \mu^-$,
together with the branching ratio and the lepton energy asymmetry,
could be a good way of defeating  the intrinsic background from
CP-conserving and indirect CP-violating amplitudes and
the experimental background from $\mu^+ \mu^- \gamma \gamma$.
The large predicted asymmetries could be measured with sufficient
statistics at new intense proton accelerators such as the Brookhaven
National Laboratory
AGS, the Fermilab main injector, or the Japanese Hadron Factory.
An examination of the functional form
of  the form factors $F_S$ and $F_V$ is needed to see if the present 
form in terms of the parameters $a_V$ and $a_S$ is adequate.
 Examination of the experimental technique
  to measure the different components of the polarization in
the laboratory as well as
the $\mu \mu \gamma \gamma$ background
is  needed to understand the possible sensitivity to 
the asymmetries.

\noindent

We  thank Laurence Littenberg, German Valencia, William Marciano,
 Steve Kettell, John Donoghue, Fabrizio Gabbiani
 for useful discussions.  This work was
supported by DOE grant DE-AC02-98CH10886.


\begin{thebibliography}{99}


\bibitem{winwolf} B. Winstein and L. Wolfenstein,
Rev. Mod. Phys. {\bf 65}, 1113 (1993).


\bibitem{rwoj} J. L. Ritchie, and S. G. Wojcicki, Rev. Mod. Phys.
{\bf 65}, 1149 (1993).

\bibitem{hsehgal}
P. Heiliger and L.M. Sehgal, Phys. Rev.  {\bf D 47}, 4920 (1993).

\bibitem{epr291} G. Ecker, A. Pich, E. de Rafael, Nucl. Phys. 
 {\bf B 291}, 692 (1987).

\bibitem{greenlee}
H. B. Greenlee, Phys. Rev.  {\bf D 42}, 3724 (1990).

\bibitem{donoghue}
J.F.Donoghue and F. Gabbiani, Phys. Rev.  {\bf  D 51}, 2187 (1995).


\bibitem{dambrosio}
G. D'Ambrosio, et al., JHEP {\bf 8}, 4 (1998).

\bibitem{bgeng91}
G. Belanger and  C. Q. Geng,  Phys. Rev.  {\bf D 43}, 140 (1991).

\bibitem{gabval}
F. Gabbiani and G. Valencia, hep-ph/0105006.




\bibitem{piggktev}
A. Alavi-Harati et al., 
Phys. Rev. Lett. {\bf 83}, 917
(1999)


\bibitem{na48ks}
R. Batley, et al. [NA48 Collaboration], CERN/SPSC 2000-002, 
Dec. 1999. 

\bibitem{anbg}
 Pankaj Agrawal, John N. Ng, G. Belanger, C.Q. Geng,
Phys. Rev. {\bf D 45}, 2383 (1992).


\bibitem{wise1}
Ming Lu, Mark B. Wise, and Martin J. Savage, Phys. Rev. {\bf D 46},
5026 (1992).

\bibitem{wise2}
Martin J. Savage, Mark B. Wise, Phys. Lett. {\bf B 250}, 151 (1990).

\bibitem{buchalla}
G. Buchalla, A.J. Buras,  Phys. Lett. {\bf B 336}, 263 (1994).

\bibitem{gourdin}
Michel Gourdin, PAR-LPTHE-93-24, May 1993.


\bibitem{bgt}
 G. Belanger, C.Q. Geng, P. Turcotte,
Nucl. Phys. {\bf B 390}, 253 (1993).

\bibitem{mpla}
M. Diwan and Hong Ma,
Int. Jour.  Mod. Phys. {\bf A 16}, 
 2449-2471 (2001).
 hep-ex-00073738.


\bibitem{pi0mumu}
G. Ecker, A. Pich, E. de Rafael, Nucl. Phys. {\bf B 303}, 665 (1988).
also see G. Ecker and A. Pich, Nucl. Phys. {\bf B 366}, 189 (1991).




\bibitem{e865pee}
R. Appel, et al., Phys. Rev. Lett. {\bf 83}, 4482 (1999).

\bibitem{e787pmm}
S. Adler, et al., Phys. Rev. Lett. {\bf 79}, 4756 (1997).

\bibitem{e865pmm}
H. Ma, et al., Phys. Rev. Lett. {\bf 84}, 2580 (2000).



\bibitem{g8}
J. Kambor, J. Missimer, and D. Wyler, Phys. Lett. {\bf B 261}, 496 (1991).

\bibitem{cohen}
A. G. Cohen, G. Ecker, A. Pich, Phys. Lett. {\bf  B 304}, 347 (1993).

\bibitem{ktevpmm}
A. Alavi-Harati et al., Phys. Rev. Lett. {\bf 84}, 5279 (2000).

\bibitem{ktevpee}
A. Alavi-Harati et al., Phys. Rev. Lett. {\bf 86}, 397 (2000).






















\end{thebibliography}
\end{document}